\documentclass[pla,reprint,onecolumn]{elsarticle}
\usepackage{amsmath,mathrsfs,amssymb,physics,braket,subfloat,graphicx,subfigure,epstopdf,lastpage,hyperref}
\hypersetup{colorlinks=true, urlcolor=red, citecolor=blue, linkcolor=blue}
\allowdisplaybreaks

\begin{document}
\title{Realistic continuous-variable quantum teleportation using a displaced Fock state channel}
\author{Deepak}
\ead{deepak20dalal19@gmail.com}

\author{Arpita Chatterjee\corref{new}}
\ead{arpita.sps@gmail.com}

\cortext[new]{Corresponding author}
\date{\today}
\address{Department of Mathematics, J. C. Bose University of Science and Technology,\\ YMCA, Faridabad 121006, India}

\begin{abstract}

We investigate ideal and non-ideal continuous-variable quantum teleportation protocols realized by using an entangled displaced Fock state resource. The characteristic function formulation is applied to measure the relative performance of displaced Fock state for teleporting squeezed and coherent states. It is found that for such single-mode input fields, the average fidelity remains at the classical threshold, suggesting that the displaced Fock states are not advantageous for teleportation. We also discuss the major decoherence effects, caused by the inaccuracy in Bell measurements and
photon losses for the propagation of optical fields via fibre channels. The changes in the teleportation fidelity are described by adjusting the gain factor ($g$), reflectivity ($R$), mode damping ($\tau$), and the number of thermal photons ($n_\mathrm{th}$). The possibility of successful teleportation can be optimized by fixing these realistic parameters.

\end{abstract}

\maketitle

\section{Introduction}

Displaced Fock state (DFS) \cite{m1,m2,m3}, a highly nonclassical quantized electromagnetic field state, owns a significant role in quantum optics, especially for representing various quasiprobability distributions. It acts as the eigenstate of the Jaynes-Cummings model consisting of coherently driven atoms in the cavity QED system \cite{alsing}. Cahill and Glauber introduced the displaced Fock state to express different quasiprobabilities in a simplified way \cite{p5,p6}. Later Boiteux and Levelut \cite{p7} suggested a specific notation for these types of states, named as semi-coherent states and investigated their basic properties. They derived a generalized completeness relation for these states. Roy and Singh \cite{p8} and Satyanarayana \cite{p9} considered these states in the framework of generalized coherent states and analyzed different properties. de Oliveira et. al. \cite{p10} proposed these states with major attention on their physical properties as amplified occupation number states and labeled them as displaced number states. C. F. Lo discussed the possibility of producing displaced and squeezed Fock states from Fock state using a general driven time-dependent oscillator \cite{lo}.  Marchiolli et. al. studied in detail the quantum interference effects induced by the superposition of DFSs in the trapped-ion system \cite{jose}. A number of optical schemes have been presented by Podoshvedov to understand the irreversible analog of the Hadamard gate while extracting displaced number states from squeezed coherent
state \cite{edns}. An experimental realization of the displaced number state is reported in \cite{ziesel}, where Ziesel et. al. first prepared Fock states ($\ket{0}$, $\ket{1}$ and $\ket{2}$) by using a laser-driven technique to lift the Jaynes-Cummings ladder, followed by a displacement (of amplitudes up to $\alpha\approx 2.8$) fabricated by means of a sudden shift in the electrostatic trapping potential. Recently, Priya et. al. \cite{malpani2019} investigated the collective effect of adding and subtracting multiple photons on the nonclassicality and phase properties of a DFS. It seems that different characters of DFS are studied extensively. However, it may be noted that a few initiatives have been taken to expose the applicability of displaced Fock state as a quantum channel in continuous-variable (CV) teleportation. In the present work, using the characteristic function formalism, we investigate how useful the displaced Fock states are for CV teleportation in ideal and realistic conditions.

Theoretically, a displaced Fock state is developed by applying the unitary displacement operator to a Fock state as $\ket{\alpha, n}\equiv D(\alpha, \alpha^*)\ket{n}$, $n=0, 1, 2, \ldots $ and $\alpha$ is the amount of displacement \cite{willi}. This procedure is equivalent of adding a non–zero complex quantity $\alpha$ to the Fock state $\ket{n}$, having $n$ number of photons and thus developing a new type of quantum state $\ket{\alpha, n}$, which comprises the coherent state $\ket{\alpha}\equiv\ket{\alpha, 0}\equiv D(\alpha, \alpha^*)\ket{0}$ as well as the Fock state $\ket{n}\equiv \ket{0, n}\equiv \frac{1}{\sqrt{n!}}a^{\dagger n}\ket{0}$ as particular cases. These states configure two types of fundamental sets: one because of their coherent state character, and next as they can act as number states in a displaced structure \cite{m1}. In the first formulation, DFSs describe an overcomplete basis (just similar to the coherent states) and in the next, the states form a complete basis (because they are orthonormal $\langle{\alpha, m}|{\alpha, n}\rangle=\delta_{m, n}$). Using these, many quantum states can be reported as an infinite superposition of displaced number states. Interest in studying the characteristics of displaced Fock state has been boosted because of their robust nonclassical features, such as quantum entanglement and negative values of the quasiprobability phase-space distributions, which are urgently needed for implementing quantum information and communication protocols \cite{dell1,kim,kita,dodo,cerf} in an efficient way and for quantum estimation tasks \cite{adesso}. In this paper, we attempt to figure out the performance of the displaced Fock state as a quantum channel for continuous-variable quantum teleportation. The input state is assumed to be either coherent or squeezed state in our approach, and the success under the Braunstein and Kimble protocol is measured in terms of the average fidelity calculated for transporting that single-mode optical state. At the same time, we try to understand the functioning of different realistic parameters in the successful teleportation of a single-mode input state through a two-mode entangled channel.

Bennett et. al. introduced the concept of quantum teleportation in the context of discrete variables \cite{a28}, and later it was realized experimentally by Bouwmeester \cite{a29} and Popescu \cite{a30}. Vaidman extended the idea of quantum teleportation in the continuous variable regime \cite{a31}. Braunstein and Kimble provided the actual quantum-optical procedure (Braunstein and Kimble (BK) protocol) for transmitting quadrature amplitudes of a light field from sender to receiver\cite{a16}, which was experimentally executed by Furusawa et. al. soon after \cite{a17,a18,a19}. In a standard teleportation protocol, an entangled resource comprising of two modes $A$ and $B$, is shared between two users, Alice and Bob, and a single-mode input state $\ket{\mathrm{in}}$ in the sender's place is the state to be teleported. The input state $\ket{\mathrm{in}}$ coupled with mode $A$ of the entangled resource by a 50:50 beam splitter, results the states $\ket{\mathrm{out}'}$ and $A'$ at the output modes of the beam-splitter. A destructive homodyne measurement is carried out by Alice on the
output modes. The results thus obtained are transmitted to Bob in a classical way; Bob implements a unitary displacement on mode $B$, which leads to the desired teleported state. This unique BK protocol has been presented in the literature in different ways \cite{van,vuk,hof}. Advancement in the theoretical description as well as the experimental implementation of CV quantum teleportation using BK protocol lies at the heart of interest of the quantum information community because of the accelerating uses of CV entangled resources in quantum information processing and quantum computing \cite{a20}. In this work, our main objective is to inspect how the performance of CV teleportation protocol with DFS resource is affected by decoherence emerging from photon losses in fibre and imperfections in Bell measurements.

The paper is arranged as follows: in Sec. \ref{sec2}, the relative performance of displaced Fock state for sending different single-mode input states in ideal condition is exhibited. In the next section, we concentrate on the effectiveness of the quantum information protocol in the presence of certain fixed values of various realistic parameters. In Sec. \ref{sec4}, we draw a conclusion based on the results obtained and discuss about some possible future extensions of the work.

\section{Ideal quantum teleportation with displaced Fock state channel}
\label{sec2}

Teleportation, an indispensable strategy for processing of quantum information, dispatches an unknown qubit from the sender to the receiver, travelling across an entangled channel shared between those two parties \cite{a28}. The measurement of teleportation can be performed in terms of the characteristic functions of the arbitrary input and resource states involved in the process \cite{marian}. The characteristic function formalism is especially helpful while managing non-Gaussian resources, because it inevitably works on the computational challenges. It is appropriate to use the fidelity of teleportation ${F}$ to evaluate the success probability of a protocol. The fidelity is a state-dependent quantity, which computes the overlap between the input state $\rho_{\mathrm{in}}$ and the teleported state $\rho_{\mathrm{out}}$, that is, ${F}=\mathrm{Tr}[\rho_{\mathrm{in}}\,\rho_{\mathrm{out}}]$. In the case of ${F} = 1$, these two states are identical, which means perfect teleportation is achieved \cite{william}. In the characteristic function description, the fidelity is defined as \cite{da1,da2}

\begin{eqnarray}
\label{eq1}
{F} = \frac{1}{\pi}\int {d^2\gamma}\,{\chi}_{\mathrm{in}}(\gamma)\,{\chi}_{\mathrm{out}}(-\gamma),
\end{eqnarray}
where
\begin{eqnarray}
{\chi}_{\mathrm{out}}(\gamma) = {\chi}_{\mathrm{in}}(\gamma){\chi}_{\mathrm{ch}}(\gamma^*, \gamma),
\end{eqnarray}
${\chi}_{\mathrm{in}}$ and ${\chi}_{\mathrm{ch}}$ are the characteristic functions of the input quantum state to be transported and the two-mode quantum channel, respectively. To calculate the average fidelity of CV teleportation, it is necessary to find out the characteristic functions of various associated input states and the two-mode entangled channel. For a single-mode optical field, the symmetric characteristic function ${\chi}(\gamma)$ can easily be derived by using ${\chi}(\gamma)=\mathrm{Tr}[D(\gamma)\rho]$, where $D(\gamma)$ is the usual displacement operator and $\rho$ is the density matrix of the related state.

Using the mathematical formula \cite{da1}
\begin{eqnarray}
\langle m|D(\alpha)|n\rangle = {\left(\frac{n!}{m!}\right)}^{1/2} {\alpha}^{m-n} e^{-\frac{1}{2}|\alpha|^2}L_n^{(m-n)}(|\alpha|^2),
\end{eqnarray}
where $L_n^{(m-n)}(|\alpha|^2)$ is the associated Laguerre polynomial, the characteristic function for the coherent state $\ket{\alpha}$ is obtained as
\begin{eqnarray}
{\chi}_{\mathrm{coh}}(\gamma)=e^{-\frac{1}{2}|\gamma|^2}e^{(\alpha^*\gamma-\alpha\gamma^*)}
\end{eqnarray}
and for squeezed state $\ket{\varepsilon}=S(\varepsilon)\ket{0}$ ($\varepsilon=r e^{i\varphi}$), it is
\begin{eqnarray}
{\chi}_{\mathrm{squ}}(\gamma)=e^{-\frac{1}{2}|\xi|^2},
\end{eqnarray}
with $\xi = \gamma \cosh{r}+\gamma^* e^{i\varphi} \sinh{r}$ \cite{da1}. A two-mode displaced Fock state is defined by \cite{ping}
\begin{eqnarray}
\ket{\alpha_1, \alpha_2, n_1, n_2} = D(\alpha_1)D(\alpha_2)\ket{n_1, n_2},
\end{eqnarray}
where $\ket{n_1, n_2} \equiv \ket{n_1}\otimes\ket{n_2}$ is a bimodal Fock state. The characteristic function for the two-mode DFS is obtained as
\begin{eqnarray}\nonumber
{\chi}_{\mathrm{DFS}}(\gamma_1, \gamma_2)&  = & \langle D(\gamma_1) D(\gamma_2) \rangle\\\nonumber
& = & e^{-\frac{1}{2}|\gamma_1|^2-\frac{1}{2}|\gamma_2|^2}e^{(\alpha_1^*\gamma_1-\alpha_1\gamma_1^*)}e^{(\alpha_2^*\gamma_2-\alpha_2\gamma_2^*)}\\
& & L_{n_1}(|\gamma_1|^2)L_{n_2}(|\gamma_2|^2)
\end{eqnarray}
An apparent way to calculate the fidelity function in \eqref{eq1} for a number of different input states and an entangled DFS resource is by applying the integration over Gaussian averages as
\begin{eqnarray*}
\int \frac{d^2z}{\pi}\exp(\xi|z|^2+\nu z+\eta z^*) = \frac{1}{\sqrt{\xi^2}}\exp(-\frac{\xi \nu \eta}{\xi^2}),
\end{eqnarray*}
[see \ref{si} for detailed calculation]

\subsection{For input coherent state $\ket{\alpha}$}

The fidelity for input coherent state is obtained as
\begin{equation}
\label{icic}
{F} = \frac{1}{\pi}\int{d^2\gamma}\,e^{-|\gamma|^2}\,e^{-(\alpha_1^*-\alpha_2)\gamma^*-(\alpha_2^*-\alpha_1)\gamma-|\gamma|^2}L_{n_1}(|\gamma|^2)L_{n_2}(|\gamma|^2)
\end{equation}

\begin{figure}[hbt]
\centering
\includegraphics[scale=.3]{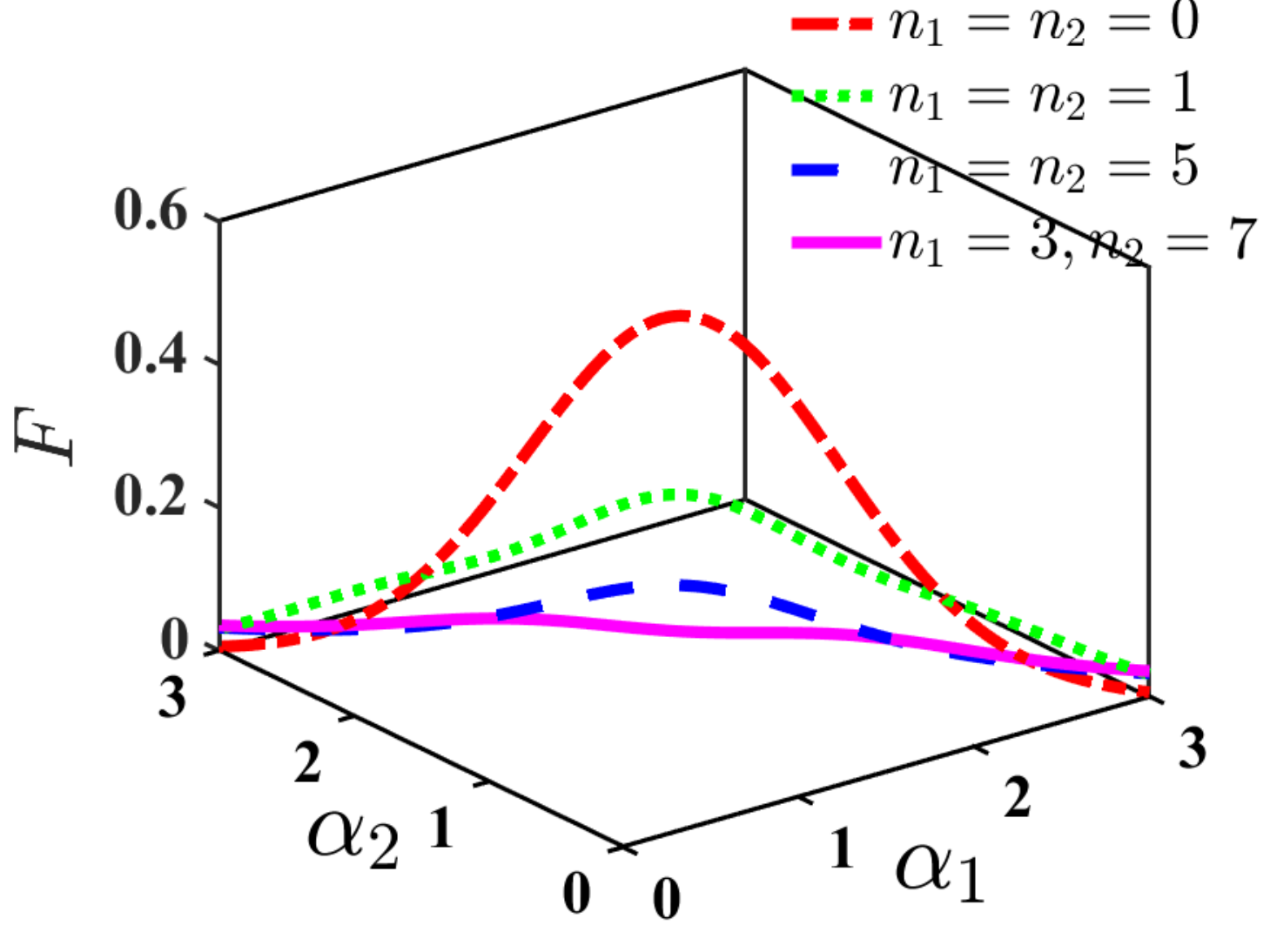}
\caption{Fidelity ${F}$ as a function of the DFS parameters $\alpha_1$ and $\alpha_2$ and for different photon numbers in mode 1 and mode 2 such as $n_1=n_2=0$ (red dash-dotted line), $n_1=n_2=1$ (green dotted line), $n_1=n_2=5$ (blue dashed line) and $n_1=3$, $n_2=7$ (magenta solid line), respectively.}
\label{fig2}
\end{figure}
In Fig.~\ref{fig2}, the dependence of fidelity on the DFS parameters $\alpha_1$ and $\alpha_2$ is shown for different $n_1$, $n_2$ values.
It is important to note that the complex amplitude $\alpha$ of the coherent state to be teleported is not arising directly in the fidelity equation. For larger values of $n_1$ and $n_2$, the fidelity decreases steadily. If $n_1\neq n_2$, the fidelity is less as compared to the fidelity attained when $n_1=n_2$. Unfortunately, the maximum fidelity obtained here is 0.499 which is below the classical benchmark. The maximum fidelity is realized at $n_1=n_2=0$, that means when the channel becomes a purely bipartite coherent state.

\subsection{For input squeezed state $\ket{\varepsilon}$}

\begin{figure}[hbt]
\centering
\includegraphics[scale=.9]{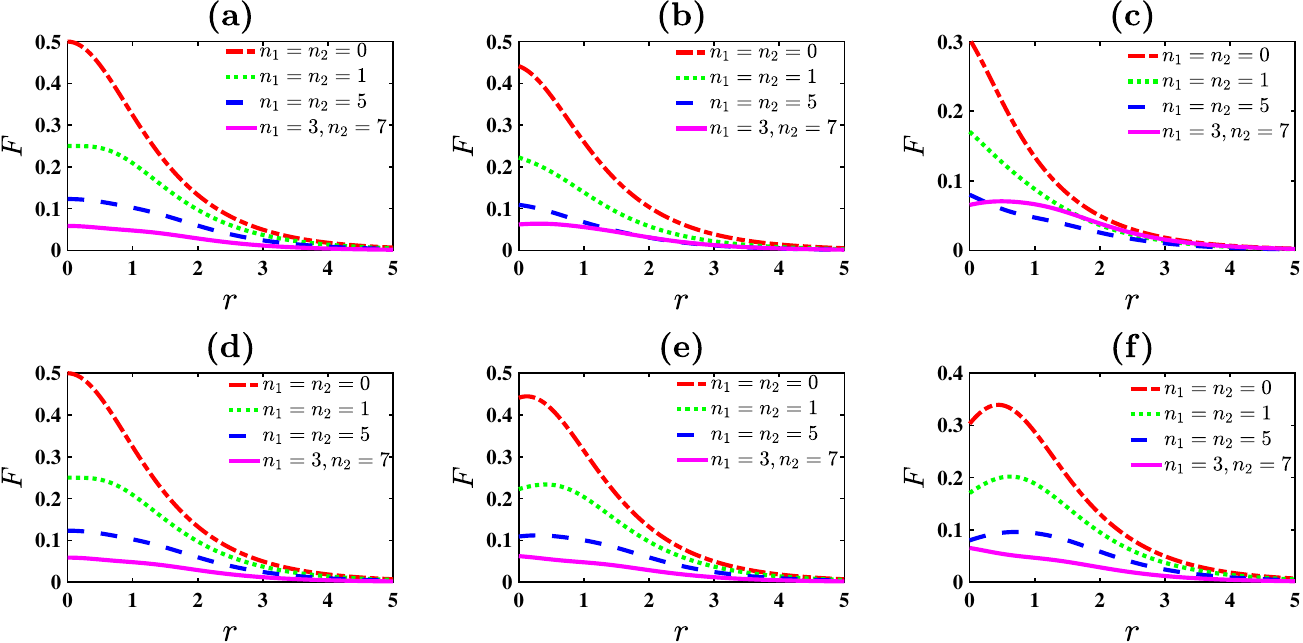}
\caption{Fidelity in ideal condition with respect to the squeezing parameter $r$ and for DFS parametric amplitudes
as (a) $\alpha_1=\alpha_2=0$, (b) $\alpha_1=0$ and $\alpha_2= 0.5$, (c) $\alpha_1=1$ and $\alpha_2=0$. In (a)-(c), the squeezing parameter $\varphi$ is 0 whereas $\varphi$ corresponds to $\pi$ in (d)-(f).}
\label{fig3}
\end{figure}
\begin{figure}[hbt]
\centering
\includegraphics[scale=.9]{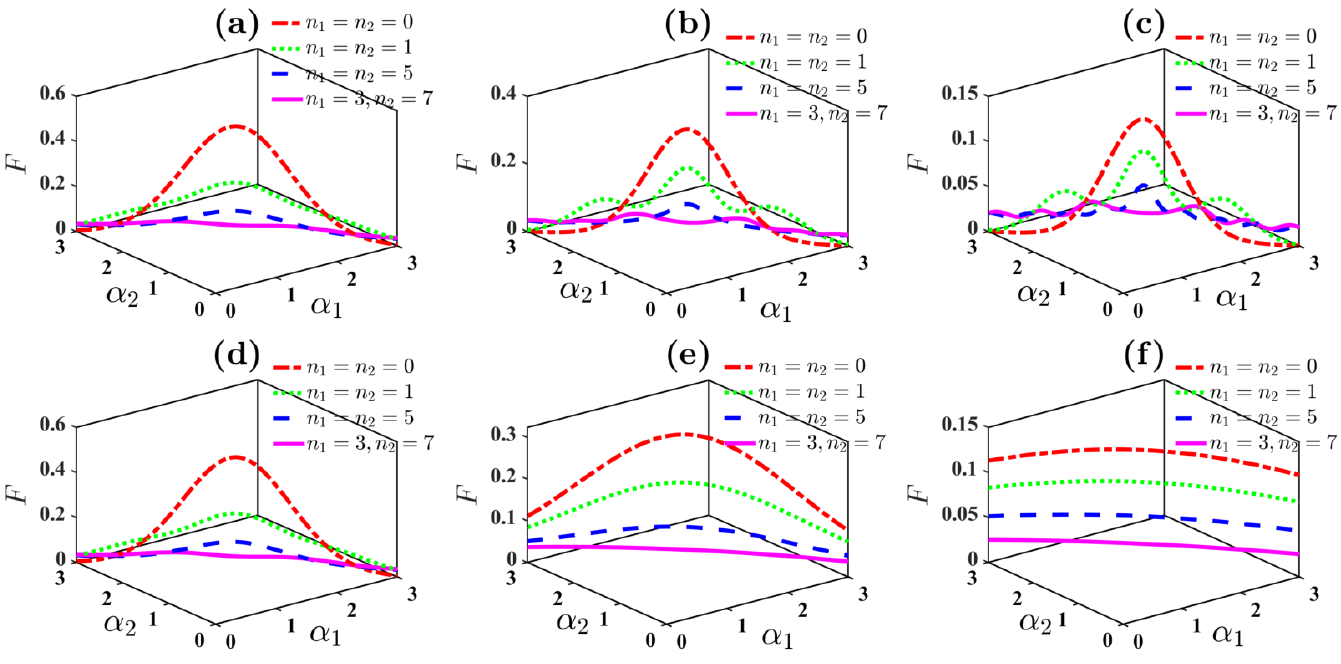}
\caption{Average fidelity (${F}$) for teleportation of a single-mode squeezed state (of different amplitudes) by using a two-mode displaced Fock state. ${F}$ is shown as a function of DFS parameters $\alpha_1$ and $\alpha_2$ and for different squeezing parameter values (a) $r=0$ and $\varphi=0$, (b) $r = 1$ and $\varphi=0$, (c) $r = 2$ and $\varphi=0$. In (d)-(f), ${F}$ corresponds to $\varphi=\pi$ and same $r$ as in (a)-(c), respectively.}
\label{fig4}
\end{figure}

The Fidelity for input squeezed state is plotted in Fig.~\ref{fig3} with respect to the squeezing parameter $r$ and for different coherent amplitudes of the two-mode DFS channel. In this case, the maximum fidelity (0.497 in Fig.~\ref{fig3}(a) and 0.499 in Fig.~\ref{fig3}(d)) values are attained when both the DFS parameters $\alpha_1$ and $\alpha_2$ are zero. For any of the $\alpha_1$ and $\alpha_2$ values, the fidelity is maximum when $n_1=n_2=0$, and it decreases steadily as $n_1$ and $n_2$ are increasing. The DFS channel seems to be more suitable for teleporting a squeezed state while $\varphi$ changes from 0 to $\pi$.

Fig.~\ref{fig4} gives the variation of ${F}$ as a function of the complex amplitudes $\alpha_1$ and $\alpha_2$ corresponding to the squeezing parameters $r = 0, 1, 2$. We note that as $r$ increases from 0 to 2, ${F}_{\mathrm{max}}$ is dropped from 0.499 to 0.132. Even for highly squeezed states ($r=2$), no enhancement in the maximum value of the average fidelity above classical limit is seen in Fig.~\ref{fig4}. Thus the displaced Fock state is not resourceful for quantum teleportation of a squeezed state.

\section{Non-ideal quantum teleportation with displaced Fock state channel}
\label{sec3}

In this section, we study the performance of the DFS resources in realistic Braunstein-Kimble CV teleportation protocol \cite{him}. The amount of success in a teleportation arrangement is strongly affected by the inefficiency of the homodyne detectors performing the Bell-measurement and the decoherence effect caused by the photon losses in noisy channel. A diagrammatic proposal for the non-ideal teleportation protocol is given in \cite{da2,leo} where at the beginning, the input and entangled resource states are considered to be pure.

For any arbitrary single-mode input state and an entangled channel, the collective effect of the passage of the field mode in a damping channel and non-ideal
Bell-measurement decides the characteristic function ${\chi}_\mathrm{out}(x_2, p_2)$ of the final output state as follows \cite{da2}:
\begin{eqnarray}\nonumber
{\chi}_\mathrm{out}(x_2, p_2) & = & {\chi}_\mathrm{in}(gT x_2, gT p_2)\,\,{\chi}_\mathrm{ch}(gT x_2, -gT p_2; e^{-\frac{\tau}{2}}x_2, e^{-\frac{\tau}{2}}p_2)\\
& & \exp[-\frac{1}{2}{\Gamma_{\tau, R}}(x_2^2+p_2^2)],
\label{eq9}
\end{eqnarray}
where $g$ is the gain factor, $\Upsilon$ is the mode damping rate, $\tau=\Upsilon t$, $\Gamma_{\tau, R}$ is the thermal phase-space covariance given by
\begin{eqnarray}
\Gamma_{\tau, R} = (1-e^{-\tau})\left(\frac{1}{2}+n_\mathrm{th}\right)+g^2 R^2
\end{eqnarray}
and $T^2$ and $R^2$ are the transmissivity and reflectivity of the beam splitter that characterizes the losses, respectively, with $T^2+R^2=1$. This equation \eqref{eq9} highlights how the teleportation protocol is affected by the two sources of noise, associated with the damping rate $\Upsilon$ and the reflectivity parameter $R^2$. A scaling factor $T$ is included in the arguments of the input characteristic function ${\chi}_\mathrm{in}$ and resource characteristic function ${\chi}_\mathrm{ch}$ to provide the effect of realistic Bell measurements. The streamlined output characteristic function in the ideal situation
\begin{eqnarray}
{\chi}_\mathrm{out}(x_2, p_2) & = & {\chi}_\mathrm{in}(x_2, p_2)\,\,{\chi}_\mathrm{ch}(x_2, -p_2; x_2, p_2),
\end{eqnarray}
can be recovered from \eqref{eq9} while substituting $R=0$ that means $T=1$, $\Upsilon=0$ i.e. $\tau=0$, and $g=1$. In the mechanism of characteristic functions, fidelity refers
\begin{eqnarray}\nonumber
{F} & = & \frac{1}{\pi} \int d^2\gamma\, {\chi}_{\mathrm{in}}(\gamma){\chi}_{\mathrm{out}}(-\gamma)\\
& = & \frac{1}{2\pi}\int dx_2 dp_2\, {\chi}_{\mathrm{in}}(x_2, p_2){\chi}_{\mathrm{out}}(-x_2, -p_2),
\end{eqnarray}
where $\gamma=\frac{1}{\sqrt{2}}(x_2+i p_2)$, $d^2\gamma=\frac{1}{2}dx_2\,dp_2$, and ${\chi}_{\mathrm{out}}(\gamma)={\chi}_{\mathrm{out}}(x_2, p_2)$ is given by \eqref{eq9}. This fidelity equation measures the performance of a displaced Fock state channel for transferring quantum information across long distance in a realistic environment.

\subsection{For input coherent state $\ket{\alpha}$}

After simple calculations, the fidelity for input coherent state in the realistic situation is as follows:

\begin{eqnarray}\nonumber
{F} & = & \frac{1}{2\pi}\int{d^2\gamma}\,e^{-(1+g^2T^2)\frac{|\gamma|^2}{4}}\,e^{(1-gT)(\alpha^*\gamma-\alpha\gamma^*)/\sqrt{2}-
\Gamma_{\tau,R}|\gamma|^2}\\\nonumber
& & \exp\left\{(gT\alpha_1-e^{-\tau/2}
\alpha_2^*)\gamma/\sqrt{2}-(gT\alpha_1^*-e^{-\tau/2}\alpha_2)\gamma^*/\sqrt{2}-(g^2T^2+e^{-\tau})/4|\,\gamma|^2\right\}\\
& & L_{n_1}\bigg(e^{-\tau}\frac{|\gamma|^2}{2}
\bigg)L_{n_2}\bigg(g^2T^2\frac{|\gamma|^2}{2}\bigg)
\end{eqnarray}

\begin{figure}[hbt]
\centering
\includegraphics[scale=.9]{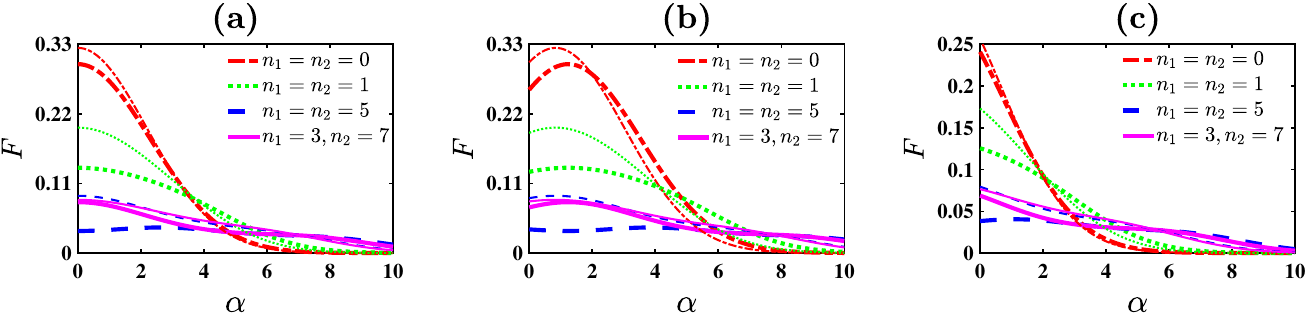}
\caption{Fidelities of realistic teleportation in terms of the input coherent state amplitude $\alpha$ with $R = 0.8$, $\tau = 0.02$, $n_{\mathrm{th}} =0$, $g = 1$, and in (a) $\alpha_1 = \alpha_2 = 0$, (b) $\alpha_1= 0$, $\alpha_2 = 0.5$ and (c) $\alpha_1 = 1$, $\alpha_2 = 0$. Here different types of lines represent the fidelities of teleporting an input coherent state $\ket{\alpha}$ corresponding to the following cases: $n_1=n_2=0$ (red dash-dotted line), $n_1=n_2=1$ (green dotted line), $n_1=n_2=5$ (blue dashed line), and $n_1=3$, $n_2=7$ (magenta solid line), respectively. The corresponding thin lines in (a) to (c) are for the fidelities with $\tau=0.8$.}
\label{fig5}
\end{figure}

\begin{figure}[hbt]
\centering
\includegraphics[scale=.9]{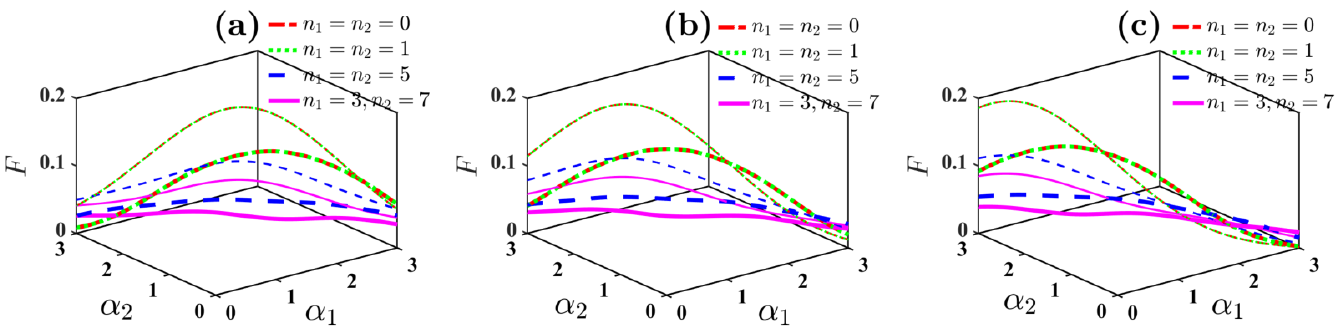}
\caption{Fidelity in realistic condition with parameters $R$, $\tau$, $n_{\mathrm{th}}$, $g$ as in Fig.~\ref{fig5} and as a function of DFS amplitudes $\alpha_1$ and $\alpha_2$, with (a) $\alpha = 0$, (b) $\alpha = 2$, and (c) $\alpha = 4$, respectively.}
\label{fig6}
\end{figure}

\begin{figure}[hbt]
\centering
\includegraphics[scale=.9]{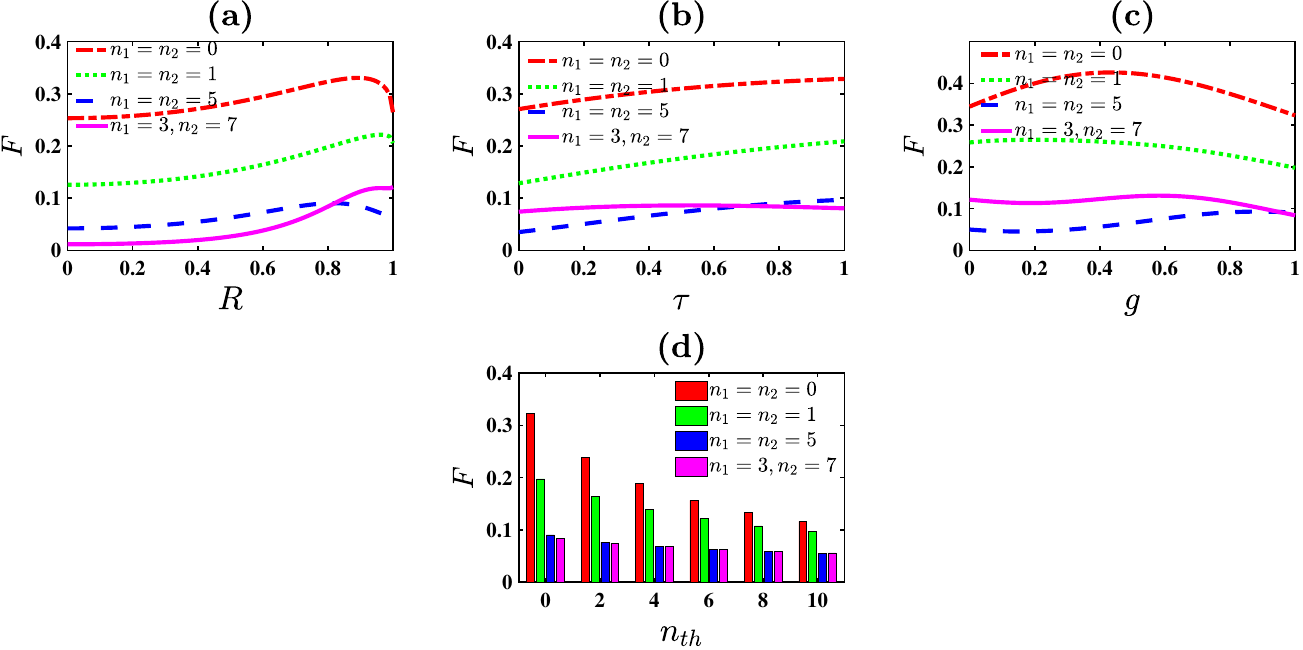}
\caption{Variation of fidelity for teleportation of a coherent state $\ket{\alpha}$ with respect to different realistic parameters, assuming $\alpha_1=\alpha_2=1$, $R=0.8$, $\tau=0.8$, $n_{\mathrm{th}}=0$ and $g=1$ whenever needed.}
\label{fig6.1}
\end{figure}

Figure~\ref{fig5} describes the variation of average fidelity for sending a coherent state $\ket{\alpha}$ across a long channel. In order to understand the simultaneous effect of different sources of decoherence in losing fidelity, the realistic parameters are fixed at $R=0.8$, $n_{\mathrm{th}}=0$ and $g=1$. The losses mainly due to the imperfections in Bell measurements (when $\tau$ is very low, $\tau=$0.02) are shown by broad lines. The corresponding thin lines exhibit the fidelity for sufficiently noisy resources ($\tau=0.8$). It is observed that the fidelity decreases monotonically with the increase of photon numbers. This behaviour is similar to the case of the ideal protocol. Interestingly, when the channel decay rate increases from 0.02  to 0.8, the DFS channel displays larger fidelity values in a certain range of $\alpha$. When $\alpha \approx 4$, the fidelity curve corresponding to $\tau=0.8$ lowers than the curve associated with $\tau=0.02$. This strange behaviour may be explained as follows: for not too large values of $\alpha$, the destructive quality of $\tau$ is reduced, while the presence of a large number of optical photons ($\geq 4$) stimulates the decoherence, and leads to a suppression of fidelity. In Fig.~\ref{fig6}, choosing the same values of $R$, $\tau$, $n_{\mathrm{th}}$ as in Fig.~\ref{fig5}, we plot ${F}$ as a function of DFS parameters $\alpha_1$ and $\alpha_2$, for fixed $\alpha$. It is clear that the disruptive effect of the noisy channel is dominating over fidelity for a large number of optical photons only, which supports our findings from Fig.~\ref{fig5}. The fidelity $F$ is plotted with respect to different realistic parameters in Fig.~\ref{fig6.1}. It is noted that increasing the inefficiency of the photodetectors has a mere effect on fidelity. The maximum fidelity is obtained when $R\approx 0.8$.


\subsection{For input squeezed state $\ket{\varepsilon}$}

The Fidelity for a squeezed state input after doing some simple calculations is derived as follows:
\begin{eqnarray}\nonumber
{F} & = & \frac{1}{2\pi}\int{d^2\gamma}\,e^{-(1+g^2T^2)\frac{|\xi|^2}{4}-\Gamma_{\tau,R}|\gamma|^2}\\\nonumber
& & \exp\left\{(gT\alpha_1-e^{-\tau/2}
\alpha_2^*)\gamma/\sqrt{2}-(gT\alpha_1^*-e^{-\tau/2}\alpha_2)\gamma^*/\sqrt{2}-(g^2T^2+e^{-\tau})/4|\,\gamma|^2\right\}\\
& & L_{n_1}\bigg(e^{-\tau}\frac{|\gamma|^2}{2}
\bigg)L_{n_2}\bigg(g^2T^2\frac{|\gamma|^2}{2}\bigg)
\end{eqnarray}

\begin{figure}[hbt]
\centering
\includegraphics[scale=.9]{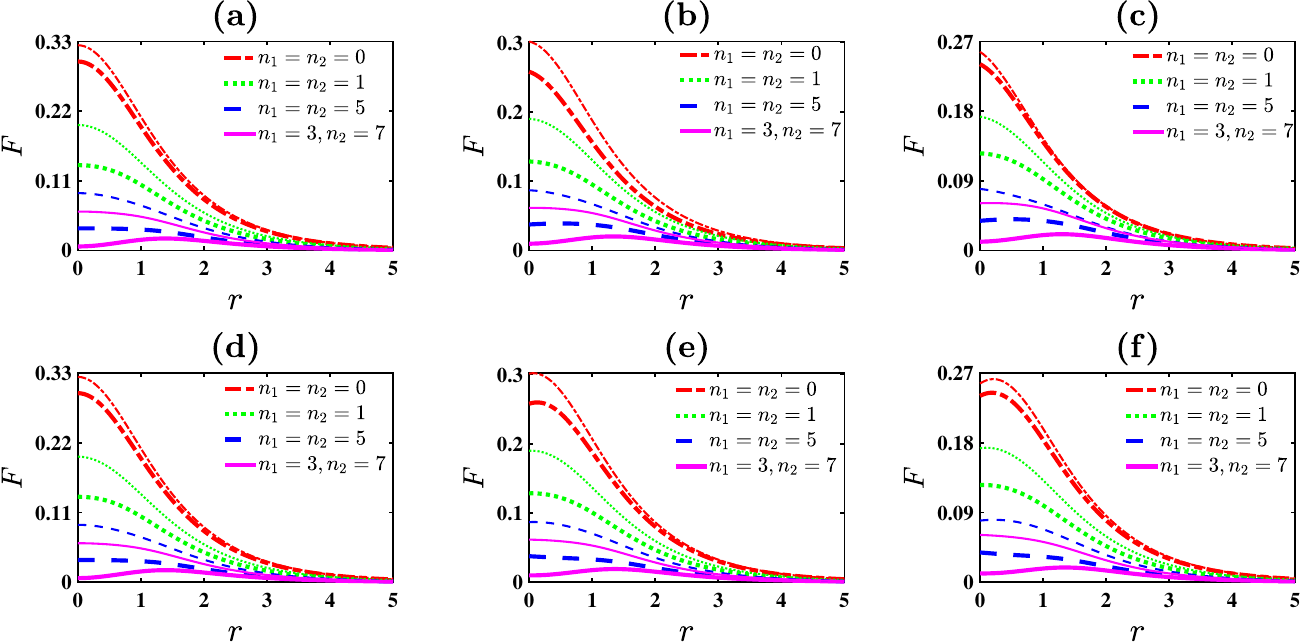}
\caption{Behaviour of fidelity in non-ideal case with respect to squeezing parameter $r$ and (a) $\varphi = 0$, $\alpha_1=\alpha_2=0$, (b) $\varphi = 0$, $\alpha_1=0$, $\alpha_2=0.5$, (c) $\varphi = 0$, $\alpha_1=1$, $\alpha_2=0$. The second row corresponds to $\varphi = \pi$ and $\alpha_1$, $\alpha_2$ are same as in (a)-(c), respectively. The plots are for different values of the realistic parameters such as $R = 0.8$, $n_{\mathrm{th}} =0$, $g = 1$ and $\tau = 0.02$ ($0.8$) corresponding to the thick (thin) lines.}
\label{fig7}
\end{figure}
\begin{figure}[hbt]
\centering
\includegraphics[scale=.9]{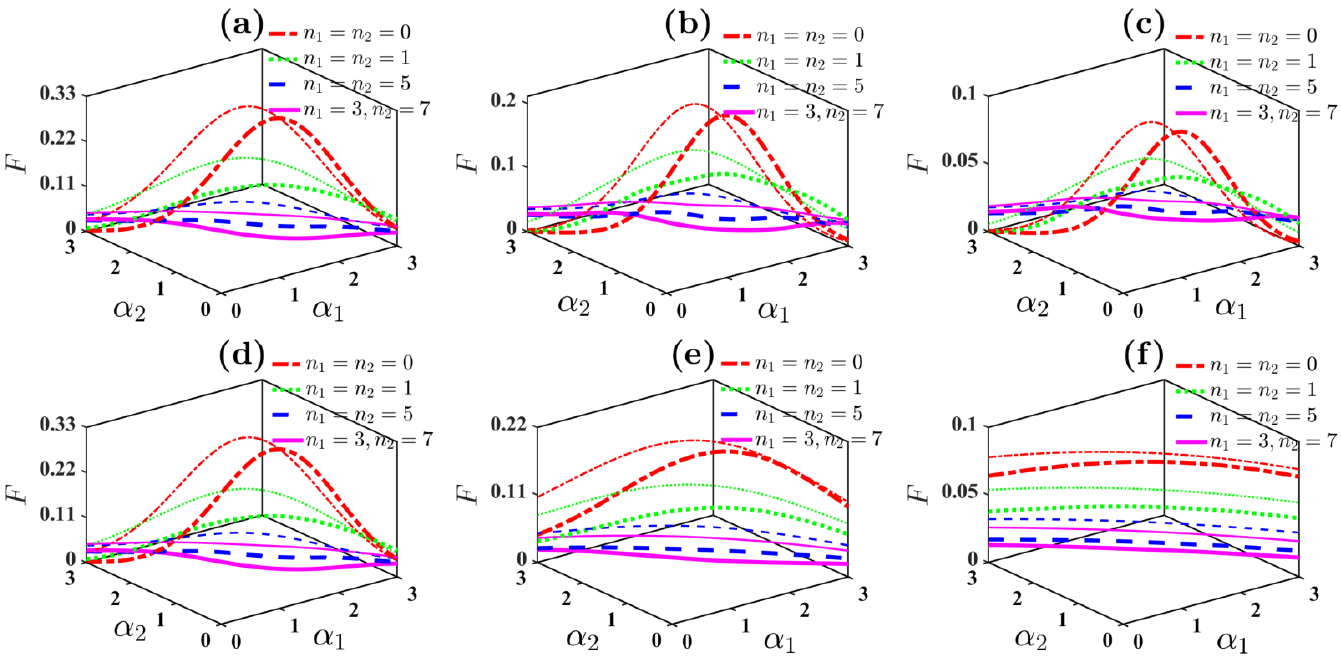}
\caption{Average fidelity as a function of DFS parameters $\alpha_1$ and $\alpha_2$, with squeezing amplitude $\varphi = 0$ and (a) $r = 0$, (b) $r = 1$, (c) $r = 2$.  In (d)-(f), ${F}$ corresponds to $\varphi = \pi$ and same $r$ values as in (a)-(c), respectively. Here $R$, $\tau$, $n_{\mathrm{th}}$ and $g$ have same values as in Fig.~\ref{fig7}.}
\label{fig8}
\end{figure}
\begin{figure}[hbt]
\centering
\includegraphics[scale=.9]{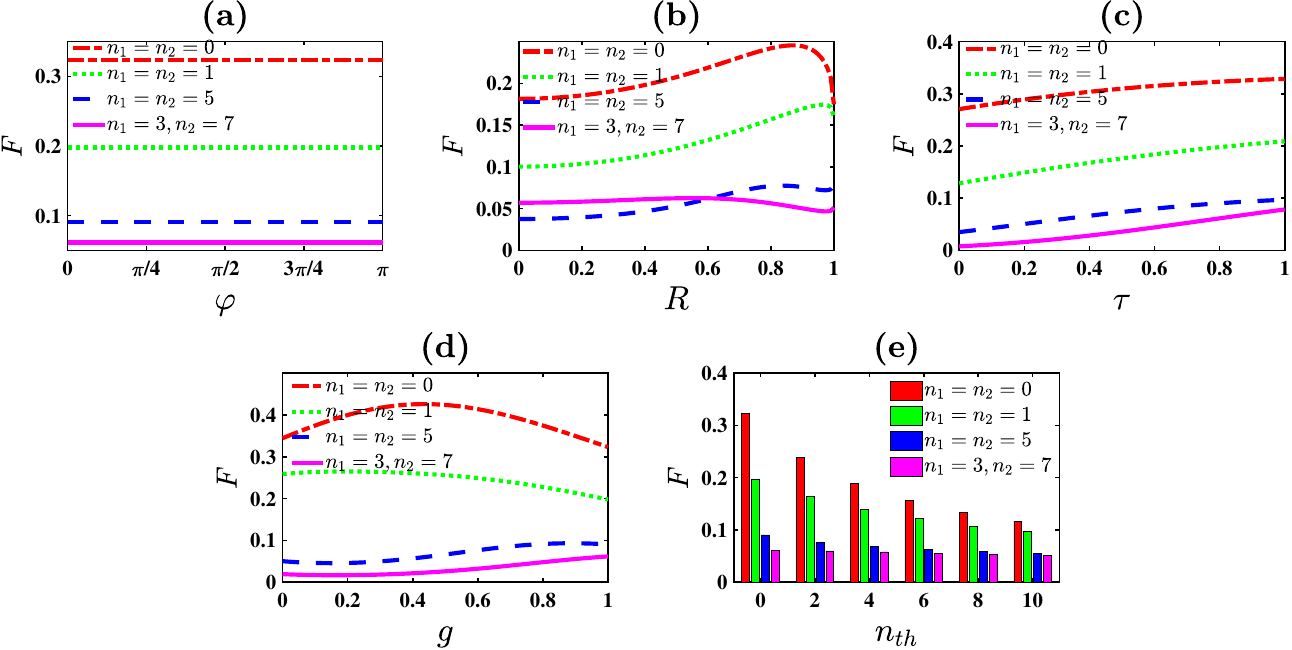}
\caption{Fidelity for input squeezed state with respect to different realistic parameters, assuming $r=0$, $\alpha_1=\alpha_2=1$, $\varphi=0$, $R=0.8$, $\tau=0.8$, $n_{\mathrm{th}}=0$ and $g=1$ in different cases.}
\label{fig8.1}
\end{figure}

In Fig.~\ref{fig7}, we continue with the same fixed parametric values as in Fig.~\ref{fig5}, i.e., $R = 0.8$, $n_{\mathrm{th}} =0$, $g = 1$, and $\tau = 0.02$ as well as 0.8, and plot the non-ideal teleportation fidelity with respect to the squeezing parameter $r$ and $\varphi=0$ (see Figs.~\ref{fig7}(a)--(c)); $r$ and $\varphi=\pi$ (see Figs.~\ref{fig7}(d)--(f)). It can be noticed that the maximum fidelity value 0.324 is obtained when $\varphi=\pi$ and squeezing is very small. Unlike the input coherent state, the decay in fidelity for increasing $r$ dominates over the degradation due to the damping $\tau$ in the entire range of $r$. Figure~\ref{fig8.1} gives the variation of fidelity with respect to different realistic parameters. Changing of only $\varphi$ has a negligible effect on fidelity. It is also seen that as the thermal photon number increases from 0 to 10, $F_{\mathrm{max}}$ decreases in a regular way.

\section{Conclusion}
\label{sec4}

In this article, the efficiency of the displaced Fock states as a quantum pathway for teleporting coherent and squeezed states are investigated. The characteristic function description is used for studying the competence of DFS in ideal and realistic scenarios. We have observed that the maximum fidelity for both the protocols is obtained when $n_1=n_2=0$. How the fidelity decoheres for transmitting via a noisy channel and imperfect homodyne detectors are analyzed. We have checked that the fidelity remains below the classical boundary which recommends that the highly-nonclassical displaced Fock states are not suitable for quantum teleportation.

For both the input states, the optimization of fidelity is performed at fixed values of $R$, $\tau$, $n_{\mathrm{th}}$, and $g$. In the context of an operational
approach, manipulating these parameters is equivalent to control different features of the experimental appliances, including the shortfall of the photodetectors and the decay rate of the noisy channel. Thus we believe that the present results will be of further use in displaced Fock state based continuous-variable quantum information processing tasks. The role of photon addition and/or subtraction for enhancing the performance of displaced Fock states in realistic teleportation will further be explored.

\begin{appendix}
\section{Details of Gaussian Integration}
\label{si}
Here we outline the description of the Gaussian integration in detail.
\begin{eqnarray}\nonumber
\label{eql}
I & = & \int{dx\,dy\,\exp(a_1x^2+a_2y^2+a_3xy+a_4x+a_5y)L_m[a(x^2+y^2)]L_n[b(x^2+y^2)]}\\\nonumber
& = & \sum_{p,q=0}^{m,n} \frac{(-a)^p(-b)^q m!n!}{p!^2q!^2(m-p)!(n-q)!}\\\nonumber
& & \times\int{dx\,dy\,\exp(a_1x^2+a_2y^2+a_3xy+a_4x+a_5y)(x^2+y^2)^{p+q}}\\\nonumber
& = & \sum_{p,q,r=0}^{m,n,p+q} \frac{(-a)^p(-b)^q m!n!(p+q)!}{p!^2q!^2(m-p)!(n-q)! r!(p+q-r)!}\\
& & \times\int{dx\,dy\,\exp(a_1x^2+a_2y^2+a_3xy+a_4x+a_5y)x^{2r}y^{2p+2q-2r}}
\end{eqnarray}
Now, rearranging the parameters as
$\mu_x=\frac{2a_2a_4-a_3a_5}{4a_1a_2-a_3^2},~~\mu_y=\frac{2a_1a_5-a_3a_4}{4a_1a_2-a_3^2},~~\rho=-\frac{a_3}{2\sqrt{a_1a_2}},~~\sigma_x=\sqrt{-\frac{2a_2}{4a_1a_2-a_3^2}},
~~\sigma_y=\sqrt{-\frac{2a_1}{4a_1a_2-a_3^2}},\,\,a_6=\frac{a_2a_4^2+a_1a_5^2-a_3a_4a_5}{4a_1a_2-a_3^2}$ and substituting $x=\sigma_xz_1+\mu_x$ and $y=\sigma_y(\rho z_1+\sqrt{1-\rho^2}z_2) +\mu_y$, ($z_1$ and $z_2$ are standard normal variate) in \eqref{eql}, we obtain \cite{gk}

\begin{eqnarray}\nonumber
I e^{a_6} & = & \sum_{p,q,r=0}^{m,n,p+q} \frac{(-a)^p(-b)^q m!n!(p+q)!}{p!^2q!^2(m-p)!(n-q)!r!(p+q-r)!}\\\nonumber
& & \times\int{dx\,dy} \exp\bigg[-\frac{1}{2(1-\rho^2)}\bigg\{\bigg(\frac{x-\mu_x}{\sigma_x}\bigg)^2+\bigg(\frac{y-\mu_y}{\sigma_y}\bigg)^2\\\nonumber
& & - \frac{2\rho}{\sigma_x\sigma_y}\bigg(\frac{x-\mu_x}{\sigma_x}\bigg)\bigg(\frac{y-\mu_y}{\sigma_y}\bigg)\bigg\}\bigg]x^{2r}y^{2p+2q-2r}\\\nonumber
& = & \sum_{p,q,r=0}^{m,n,p+q} \frac{(-a)^p(-b)^qm!n!(p+q)!}{p!^2q!^2(m-p)!(n-q)!r!(p+q-r)!}\sqrt{1-\rho^2}\sigma_x\sigma_y\\\nonumber
& & \times\sum_{s,t,u=0}^{2r,2p+2q-2r,t}{2r\choose s}{2p+2q-2r\choose t}{t\choose u}\mu_x^{2r-s}\mu_y^{2p+2q-2r-t}\\\nonumber
& & \times\sigma_x^{s}\sigma_y^{t}\rho^{t-u}(1-\rho^2)^{u/2}\int e^{-\frac{1}{2}(z_1^2+z_2^2)}z_1^{s+t-u}z_2^{u}dz_1dz_2\\\nonumber
& = &\left\{
\begin{array}{lcl}
\sum_{p,q,r=0}^{m,n,p+q} \frac{(-a)^p(-b)^qm!n!{{p+q}\choose r}}{p!^2q!^2(m-p)!(n-q)!}\sqrt{1-\rho^2}\sigma_x\sigma_y\sum_{s,t,u=0}^{2r,2p+2q-2r,t}{2r\choose s}{2p+2q-2r\choose t}{t\choose u}\\
\times\mu_x^{2r-s}\mu_y^{2p+2q-2r-t}\sigma_x^{s}\sigma_y^{t}\rho^{t-u}(1-\rho^2)^{u/2}2^{\frac{s+t}{2}+1}\,
\Gamma\left(\frac{s+t-u+1}{2}\right)\Gamma\left(\frac{u+1}{2}\right),\\\nonumber
\text{if both $s+t$ and $u$ are even}\\\\
0,\,\,\,\,\, \text{otherwise}
\end{array}
\right.
\end{eqnarray}
by using \cite{gb}
\begin{eqnarray}
\int{dx\,e^{-ax^b}x^n} =
\left\{
\begin{array}{lll}
\frac{2\,\Gamma[(n+1)/b]}{b\,a^\frac{n+1}{b}},\,\,\,\text{if $n$ is even}\\\\
0, \,\,\,\,\,\,\,\,\,\,\,\,\,\,\,\,\,\,\,\,\,\,\,\,\,\,\,\,\,\text{otherwise}
\end{array}
\right.
\end{eqnarray}

\end{appendix}

\begin{center}
\textbf{ACKNOWLEDGEMENT}
\end{center}

Deepak acknowledges support from the Council of Scientific and Industrial Research
(CSIR), Govt. of India (Award no. 09/1256(0006)/2019-EMR-1).

\begin{center}
\textbf{DATA AVAILABILITY STATEMENT}
\end{center}

Data will be made available on reasonable request.

\end{document}